\documentclass[sort&compress]{elsarticle}
\pdfoutput=1
   
\usepackage{subfloat}
\usepackage{subfig}
\usepackage{arydshln}
\usepackage{multirow}
\usepackage{url}
\usepackage{amssymb}
\usepackage{times}
\usepackage{ifthen}
 \usepackage{setspace}

\usepackage{amssymb}

\newenvironment{proof}{\textsc{Proof.}}{~$\square$\\ }
\usepackage{graphicx}
\graphicspath{{./plots/}} 
\usepackage{pslatex}

\makeatletter

\usepackage[normalem]{ulem}
 \usepackage{verbatim}
\usepackage{pslatex}
\usepackage{amsmath}
\usepackage{ntheorem}
 
\newtheorem{proposition}{Proposition} 
\newtheorem{lemma}{Lemma} 
\newtheorem{corollary}{Corollary}

\theoremheaderfont{\sc}\theorembodyfont{\upshape} 
\theoremstyle{nonumberplain} 
\theoremseparator{} 
\theoremsymbol{\rule{1ex}{1ex}} 
\usepackage{color}
\usepackage{hyperref}

\usepackage{ifpdf}

\usepackage{algorithmic}
\usepackage{algorithm}

\begin{document}

\begin{frontmatter}

\title{The universality of iterated hashing over variable-length strings}

 \author[UQAM]{Daniel Lemire\corref{cor1}} \ead{lemire@acm.org} 

 \address[UQAM]{\scriptsize LICEF, Universit\'e du Qu\'ebec \`a Montr\'eal (UQAM), 100 Sherbrooke West, Montreal, QC, H2X 3P2 Canada
}

 \cortext[cor1]{Corresponding author. Tel.: 00+1+514 987-3000 ext. 2835; fax: 00+1+514 843-2160.}

\begin{abstract}
Iterated hash functions process strings recursively, one character at a time.
At each iteration, they compute a new hash value from the preceding hash value 
and the next character. 
We prove that iterated hashing can be pairwise independent, but never 3-wise independent.
We show that it can be almost universal over strings much longer
than the number of hash values; we bound the maximal string length given the collision probability.
\end{abstract}


\end{frontmatter}

\section{Introduction}

We consider hash functions mapping variable-length strings to $L$-bit integers.
They have numerous applications from indexing---e.g., with hash tables~\cite{KnuthV3E3,pagh2004cuckoo,byers2003simple,1496842} and Bloom filters~\cite{67941}---to spell-checking~\cite{kukich1992techniques}, compression~\cite{758900} and cryptography~\cite{rogaway1999bucket}.

We consider hash functions $h$ picked
randomly from a family $\mathcal{H}$~\cite{carter1979uch}.
We focus on \emph{iterated hash functions}~\cite{preneel1994hash,liskov2007constructing}: 
given a string $s_1 s_2 \cdots s_n$, 
starting from an initial value (or seed) $H_0$, the hash value of the whole string, $H_n$, is computed recursively from a \emph{compression function} $F$ as 
$H_{i} = F(H_{i-1},s_i)$ for $i=1,2,\ldots, n$. Thus, a hash function is defined both by an initial value $H_0$ and a compression function $F$.
A typical example is Carter-Wegman Polynomial Hashing over finite fields~\cite{krovetz2001fast,carter1979uch}, that is, hash functions of the form $h(s)=\sum_i t^{n-i} s_i$ for some
randomly chosen element $t$.
Many hash functions over variable-length strings are
iterated including 
 Pearson hashing~\cite{78978,338597},   SAX and SXX~\cite{ramakrishna1997performance} as well  as the hash functions commonly used in C++ and Java.
In cryptography, iterated hashing is also known as Merkle-Damg\aa rd~\cite{909000} hashing; it includes the popular functions MD4, MD5, SHA-0 and SHA-1. 

Good hash functions are such that hash values \emph{appear} random.
Formally, a family  is 
 (pairwise) \textit{universal} if
the probability of a collision is no larger than if the hash values were random: $P\left (h(s)=h(s')\right )\leq 1/2^L$
for $s \neq s'$.
It is $\epsilon$-almost universal~\cite{188765} (or $\epsilon$-AU) if the probability of a collision is bounded by $\epsilon<1$. 
Furthermore, a family is $k$-wise independent if given  $k$~distinct elements $s^{(1)}, s^{(2)}, \dots, s^{(k)}$, their hash
values are independent:
\begin{eqnarray*}
P\left (h(s^{(1)})= y^{(1)}  \; \land  \; h(s^{(2)})= y^{(2)}  \; \land \dots \land  \; h(s^{(k)})= y^{(k)} \right ) =\frac{1}{2^{kL}}
\end{eqnarray*}
for any hash values $y^{(1)}, y^{(2)}, \dots, y^{(k)}$.
Pairwise (or 2-wise) independence implies pairwise universality and thus, it is also called  strong universality.

The main contributions  are as follows.
\begin{itemize}
\item We show that iterated hashing cannot be 3-wise independent (\S~\ref{sec:pairwiseatbest}). Thus, we have
to be satisfied with pairwise independence. We can get 3-wise independence if we consider a generalization
of iterated hashing where there is a new compression function with each iteration in the computation.
However, 4-wise independence remains impossible.
\item We show that pairwise independence is possible for iterated hashing by presenting the \textsc{Tabulated} family (\S~\ref{sec:iteratedbytabulation}).
\item We show
that almost universality is possible for strings longer than $2^L$~characters, e.g., with the
Pearson family~(\S~\ref{sec:pearson}).  
\item Iterated hashing families have limited cardinality: there are only so many 
possible compression functions. This limits their universality.
We apply results from Nguyen and Roscoe~\cite{Ros128} and Stinson~\cite{188765} to derive new bounds (\S~\ref{sec:nguyenroscoe}).
To make one such bound tighter, we use the fact that pairwise independent families must
have permuting compression functions, a concept we introduce in \S~\ref{sec:compressionfunctions}.
\item We can derive tighter bounds using the  innate limitations of iterated hashing (\S~\ref{sec:limitations}). 
Table~\ref{fig:mainresults} summarizes some of our results.
\end{itemize}

\begin{table}[tb]
\caption{\label{fig:mainresults} Upper bounds on the string length for arbitrarily large iterated hash families. We write
$LCM_{2^L}$ for the least common multiple of the integers from 1 to $2^L$. }
\centering
\begin{tabular}{ccc}\hline
 & Cardinality-based bounds (\S~\ref{sec:nguyenroscoe})  & New bounds (\S~\ref{sec:limitations})\\\hline
\rule{0cm}{1.2em}  strongly universal & $L + 2 \log 2^L! - \log(2^L-1)-1$ & $2^L+1$  \\
\rule{0cm}{1.2em}  universal & $2 L+L 2^{L+1}$  &  $2^L+1$  \\
\rule{0cm}{1.2em} almost universal (any $\epsilon<1$) & $L( 2^{L (2^{L+1} +1)} +1)$ & $2^L+ LCM_{2^L}-1$ \\\hline
\end{tabular}
\end{table}

\section{Preliminaries}

We want $L$-bit integer hash values: hash functions
map elements to integers in $\{0,1,\ldots, 2^L-1\}$. 
The weakest property we require from a family is uniformity: all hash
values are equiprobable. That is  a family is 
uniform  if $P(h(s)=y)=\frac{1}{2^L}$ for any $s$ and $y$.
(We pick $h$ uniformly at random from the family $\mathcal{H}$.)
 It is
not difficult to construct such a family. For example, let $\mathcal{H}$
be the set of all $2^L$~distinct constant functions ($h(s)=c$ for all $s$).
This family is uniform but a poor choice in practice because any two elements $s$ and $s'$ are 
sure to collide: $P\left (h(s)=h(s') \right )=1$.

Thus, we commonly seek families satisfying stronger conditions. A family is $\epsilon$-almost universal 
if the probability of a collision is bounded by $\epsilon<1$: 
$P\left (h(s)=h(s')\right )\leq \epsilon$ for any $s$ and $s'$.
We say that the family is universal when it is $1/2^L$-almost universal.

While bounding the probability of a collision is sufficient for
some applications like conventional hash tables, other results
require stronger properties.
A family is pairwise independent (or strongly universal)
if the hash values of any two elements are independent:
\begin{eqnarray*}P\left (h(s)=y \land h(s')=y'\right )=\frac{1}{2^{2L}}\end{eqnarray*}
for any two distinct elements $s,s'$ and any two hash values $y,y'$. It is
3-wise independent if the hash values of any three elements
are independent: 
\begin{eqnarray*} P\left (h(s)=y \land h(s')=y' \land h(s'')=y''\right)=\frac{1}{2^{3L}}\end{eqnarray*} 
for any three distinct elements $s,s',s''$ and any three hash values $y,y',y''$.
We can generalize this definition to $k$-wise independence. 
A family which is $k$-wise independent for any $k$ is fully
independent. 
(A family is trivially $k$-wise independent over a set containing
less than $k$~distinct elements. In our work,
we implicitly assume that there are at least $k$~distinct elements  whenever we consider $k$-wise independence.)

We have that $k$-wise independence implies $k-1$-wise independence for
$k\geq 2$. For example, suppose we have 3-wise independence.
Then we have that
\begin{eqnarray*} 
P\left (h(s)=y \land h(s')=y' \right ) & =  & \sum_{y''=0}^{2^L-1}
P\left(h(s)=y \land h(s')=y' \land h(s'')=y''\right )\\ 
& = & 2^L \times \frac{1}{2^{3L}}=\frac{1}{2^{2L}}.
\end{eqnarray*}
Similarly, $k$-wise independence for any $k>1$ implies uniformity.

 \begin{figure}[tb]
\centering \includegraphics[width=0.5\textwidth]{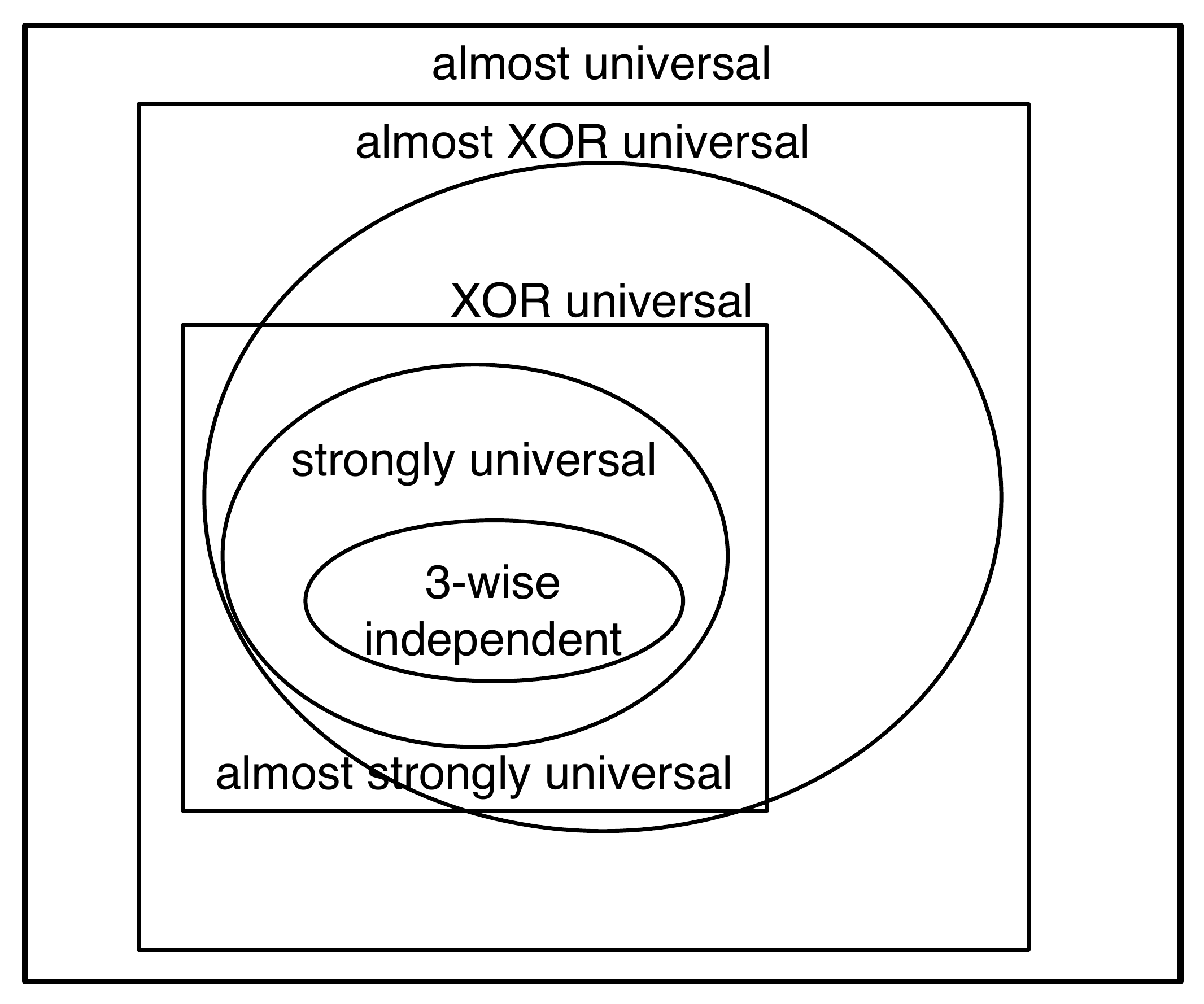}
 \caption{\label{fig:propdiagram} Visual summary of the main properties related to universality.
 Strong universality is synonymous with pairwise independence.}
 \end{figure}

A family that is universal, but not strongly universal might be  \textit{XOR universal}
 if the bitwise exclusive OR of hash values appears
random, that is  $P(h(s) \oplus h(s')=y)=1/2^L$ for all distinct
elements $s,s'$ and hash values $y$~\cite{Ros128,krawczyk1995new}.
 (The symbol $\oplus$ is the bitwise exclusive OR\@.)

We can weaken both strong universality and XOR universality:
\begin{itemize}
\item Given $\epsilon <1$, a family is
$\epsilon$-almost strongly universal if
it is uniform and if 
$P(h(s)=y \land h(s')=y')\leq \epsilon / 2^L$
for any two distinct elements $s,s'$ and any two hash values $y,y'$.
A $1/2^L$-almost strongly universal family is strongly universal.
\item Similarly, it is $\epsilon$-almost
XOR universal (or $\epsilon$-AXU) if the probability $P(h(s) \oplus h(s')=y)$ is bounded by $\epsilon$. 
 \end{itemize}
We have
that $\epsilon$-almost
strong universality implies $\epsilon$-almost XOR
 universality which itself implies  $\epsilon$-almost 
 universality.
(See Fig.\ref{fig:propdiagram}.)

We can also generalize almost universality
to $k$-wise almost universality:
a family is $k$-wise $\epsilon$-almost universal~\cite{Ros128} if
the probability of a $k$-way collision is bounded by $\epsilon$
for some $\epsilon <1$.
That is, if we have
 $P(h(s^{(1)})=h(s^{(2)})=\ldots=h(s^{(k)}) )\leq \epsilon$
 as long as the $k$~elements $s^{(1)}, s^{(2)}, \ldots, s^{(k)}$ are distinct. 
 We have that 
 $k$-wise $\epsilon$-almost universality implies
 $k+1$-wise $\epsilon$-almost universality. E.g.,
 almost universality implies 3-wise almost universality.

\section{Iterated hashing is pairwise independent at best}
\label{sec:pairwiseatbest}
We write the concatenation \texttt{ab} of two strings \texttt{a} and  \texttt{b} as 
 $\texttt{a} \parallel \texttt{b}=\texttt{ab}$.
If $\emptyset$ is the empty string, then $\emptyset  \parallel  \texttt{a}= \texttt{a}$.
We begin by characterizing iterated functions.

\begin{proposition}\label{prop:equiv}
Consider hash functions over all strings, including the empty string.
The following statements are equivalent:
\begin{itemize}
\item $\mathcal{H}$ is a family of iterated hash functions;
\item For any $h\in \mathcal{H}$, whenever 
$h(s)=h(s')$ for a pair
of strings $s,s'$, then  
$h(s \parallel  s'')=h(s'  \parallel  s'')$ for any string $s''$.
\end{itemize}
\end{proposition}
\begin{proof} By induction, iterated hash functions satisfy the second point.
Indeed, suppose that $h\in \mathcal{H}$ is an iterated function
with a corresponding compression function $F$.
Let $s''_i$ be the $i^{\mathrm{th}}$~character of the string $s''_i$. By
appending the first character of $s''$ to both strings ($s$ and $s'$) we
get a collision: 
$h(s \parallel  s''_1) = F(h(s), s''_1)= F(h(s'), s''_1) =h(s' \parallel  s''_1)$.
We can then append the remaining characters of $s''$ one by one starting with $s''_2$ and
finally prove that 
$h(s \parallel  s'')=h(s'  \parallel  s'')$.

Conversely, suppose the second point is true. Pick $h\in \mathcal{H}$.
We want to construct a corresponding compression function $F$.
For any hash value $y$ in the domain of $h$, there is at least one string $\rho_y$ such that $h(\rho_y)=y$.
Let 
$F(y,\texttt{a})=h(\rho_y \parallel \texttt{a})$ for all characters $\texttt{a}$.
By the second point, $F$ is well defined: its definition is independent of the
choice of $\rho_y$.
We can verify that the iterated hash function with compression
$F$ and initial value $H_0=h(\emptyset)$ agrees with $h$ on all strings which concludes
the proof.
\end{proof}

Families of iterated hash functions have limited independence.
The next lemma shows that they are 
pairwise independent at best. Moreover, almost strong universality (and thus strong universality)
requires a non-fixed initial value.
\begin{lemma}Iterated hashing cannot be 3-wise independent, unless we bound the string length to two characters. Moreover, almost 
strong universality  is impossible with a fixed initial value unless we bound the string length to one character.
\end{lemma}
\begin{proof}
We prove the first statement by contradiction. Suppose that an iterated family $ \mathcal{H}$
is 3-wise independent. By definition,
we must have 
\begin{eqnarray*}P\left ( h(\texttt{a})=y \land h(\texttt{ab})=y \land h(\texttt{abb})=y\right ) = \frac{1}{2^{3L}}\end{eqnarray*}
for any hash values $y$, and any characters \texttt{a} and \texttt{b}. (We allow  $\texttt{a} = \texttt{b}$.)
However, the family is also pairwise independent so that
$P\left ( h(\texttt{a})=y \land h(\texttt{ab})=y \right )= \frac{1}{2^{2L}}$.
However,  if $h(\texttt{a})=y$ and $h(\texttt{ab})=y$ then the compression function satisfies $F(y,\texttt{b})=y$ and therefore $h(\texttt{abb})=y$. Hence we conclude that 
\begin{eqnarray*}\frac{1}{2^{2L}}  & = & P\left ( h(\texttt{a})=y \land h(\texttt{ab})=y \right ) \\ & = & P\left ( h(\texttt{a})=y \land h(\texttt{ab})=y \land h(\texttt{abb})=y\right )\\ & = & \frac{1}{2^{3L}},\end{eqnarray*}
a contradiction when $L\geq 1$.

For the second statement, suppose that the family is $\epsilon$-almost universal for $\epsilon<1$.
Let the fixed initial value be $H_0$.
If the family is pairwise independent then 
$P(h( \texttt{a})=H_0 \land h( \texttt{aa})=H_0)\leq \epsilon/2^{L}$. Moreover,
because almost strong universality implies uniformity, we have that 
$P(h( \texttt{a})=H_0) = 1/2^L$.
 Because $h$ is iterated,
we 
have that $h( \texttt{a})=H_0$ implies that the compression function satisfies $F(H_0,\texttt{a})=H_0$.
Hence we have that $h( \texttt{a})=H_0$ implies $h( \texttt{aa})=H_0$.
It follows that $P(h( \texttt{a})=H_0 \land h( \texttt{aa})=H_0) =P(h( \texttt{a})=H_0) = 1/2^L$ and
therefore the family cannot be pairwise independent because $ 1/2^L > \epsilon/2^{L} $.
\end{proof}

To allow better independence, we consider \emph{generalized iterated hash functions} where a new 
compression function is used for each new character:  $H_{i} = F_i(H_{i-1},s_{i})$ for $i=1,2,\ldots, n$. 
A family
of generalized hash functions is such that whenever $h(s)=h(s')$ for a pair
of strings $s,s'$ having the same length, then  
$h(s\parallel s'')=h(s'\parallel s'')$ for any string $s''$. 

It includes hashing by multilinear functions over finite fields~\cite{carter1979uch,Sarkar2008}, e.g., hash functions of the form $h(s)=m_1+\sum_i m_{i+1} s_i$
with randomly generated values $m_1, m_2, \dots$ (henceforth \textsc{Multilinear}). The compression functions are $F_i(y,c)= y + m_{i+1} c$ with an 
initial value of $m_1$.
The computation is in the finite field $\mathbb{F}_p$: characters are mapped to elements of $\mathbb{F}_p$.
For example, we can choose the field of cardinality $p=2^L$: 
the polynomials with binary coefficients modulo $p(x)$---where $p(x)$ is an irreducible polynomial of degree $L$. 
We write $\mathbb{F}_{2^L}=\textrm{GF}(2)[x]/p(x)$.
That is, integers in $[0,2^L)$ are represented as polynomials of degree $L-1$ having binary coefficients.
Addition or subtraction is the bitwise (or term-wise) exclusive OR\@.
Multiplication by $x$ is just the left shift, unless the left-most bit is 1, in which case the left shift must be followed by the addition  with $p(x)$. 
Exhaustive lists of irreducible polynomials
are available online~\cite{COS}. 
Otherwise, when $p$ is a prime number, we merely have to compute $F_i(y,c)= m_{i+1} y+c \bmod{p}$ using the usual integer algebra. 
We might prefer finite fields that have prime cardinality close to $2^L$. For example, we can set $p$ to some Mersenne prime such as
  $2^{17}-1$, $2^{31}-1$ or $2^{61}-1$, or other convenient prime such as $2^{32}-5$ or $2^{64}-$59~\cite{krovetz2001fast}. 

\begin{lemma} \textsc{Multilinear} is
pairwise independent if we forbid strings ending with the value zero.
\end{lemma}
\begin{proof}
We have that $h(\texttt{a}0)=h(\texttt{a})$ so universality is impossible if we allow strings to end with the value zero. So let $s$, $s'$ be two distinct strings of
lengths $|s|$ and $|s'|$
ending with non-zero values. Assume without loss of generality that $|s|\geq |s'|$.
Given $h(s')=y'$, we can solve for $m_1$ as a function of $y'$, $s'$ and the values $m_2, m_3, \dots, m_{|s'|}$.

If $s$ is longer than $s'$ (that is $|s|>|s'|$), then 
 we can solve for $m_{|s|}$ in $h(s)=y$ as a function of $y$, $s$ and $m_1, m_3, \dots, m_{|s|-1}$. In turn, if we substitute our
 solution for $m_1$, we have $m_{|s|}$ as a function of $y$, $y'$, $s$, $s'$, and all $m_i$ for $i\neq 1, |s|$.
 
 If $s=s'$, then there must be some $j$ such that $s_j\neq s'_j$. Hence, we can solve for $m_j$ in $h(s)-h(s')=y-y'$ as a function of $y$, $y'$, $s$, $s'$ and all
 $m_i$ for $i\neq 1, j$ after substituting the solution for $m_1$ from  $h(s')=y'$.
 
 Thus, in either case, among all possibles values of $m_1, m_2, \dots, m_{|s|}$, two values are fixed by $h(s)=y \; \land \; h(s')=y' $. Hence, we have
 that $P(h(s)=y \; \land \; h(s')= y')= p^{|s|-2}/p^{|s|}=1/p^2$ where $p$ is the cardinality of the field. This proves pairwise independence. 
\end{proof}

If we choose zero as an initial value $m_1=0$, then  \textsc{Multilinear} is still XOR universal when $p=2^L$. However, it fails to be  pairwise independent.
Indeed, given \texttt{a}, \texttt{b} two distinct elements of $\mathbb{F}_p$, we cannot satisfy both $h(\texttt{a})=m_1 + m_2 \texttt{a} =\texttt{a}$ and $h(\texttt{b})= m_1 + m_2 \texttt{b} =\texttt{a}$ unless $m_1\neq 0$.

\textsc{Multilinear} has a nearly optimal memory-universality trade-off. Indeed, Stinson~\cite{188765} showed
that pairwise independent families must have cardinality at least $1+a(b-1)$ where $a$ is the number of strings
and $b$ is the number of hash values. There are $p^n-1$~strings of length at most $n$ in $\mathbb{F}_p$ ending with a non-zero value.
Thus, any pairwise independent family from strings  in  $\mathbb{F}_p$ of length  bounded by $n$ to elements in  $\mathbb{F}_p$ must
have at least $1+(p^n-1) (p-1)$~hash functions. That is, its cardinality is in $\Omega(p^{n+1})$. Meanwhile, there are $p^{n+1}$~different hash functions in \textsc{Multilinear} when strings have length at most $n$.

We can have better universality than \textsc{Multilinear}, at the cost of a higher memory usage. 
Consider sequences of 3-wise independent hash functions $h_i$ from characters to $L$-bit integers. 
The Zobrist~\cite{zobrist1970,zobrist1990new} family of string hash functions
$h(s) = h_1 (s_1) \oplus h_2 (s_2)\oplus \dots \oplus h_n(s_n)$ is
3-wise independent~\cite{lemi:one-pass-journal,thorup2004tabulation,carter1979uch}. It is also an example of
 generalized iterated hashing with the compression function $F_i(y,c)=y\oplus h_i( c )$.
  Moreover, it has optimal independence
  since generalized iterated families are 3-wise independent at best according to the next lemma.

\begin{lemma}Generalized iterated hashing cannot be 4-wise independent unless we bound the string length to one character.
\end{lemma}
\begin{proof}
Consider any generalized iterated hash function $h$. If $h(s)=h(s')$ for any two 
strings $s$ and $s'$ of the same length then $h(s\parallel \texttt{a})=h(s'\parallel \texttt{a})$ for any character \texttt{a}.
Hence, assuming that the family is 4-wise independent, we have that
\begin{eqnarray*}
\frac{1}{2^{4L}} & = & P\left (h(s)=y \land h(s')= y \land h(s\parallel \texttt{a}) = z  \land h(s'\parallel \texttt{a}) =z' \right )\\
                 & = & 0
\end{eqnarray*}
whenever $z \neq z'$, a contradiction.
\end{proof}

In the following sections, we consider solely conventional iterated hashing.

\section{Pairwise independence requires permuting compression functions}
\label{sec:compressionfunctions}
Permuting compression functions $F$ are such that $y \mapsto F(y,c)$ is a permutation of the hash values $y\in[0,2^L)$
for any character $c$ and any compression function. Hence, if $y\neq y'$ then $F(y,c) \neq F(y',c)$ when $F$ is permuting.

(It is not necessary for $F$ to permute all integer values in $[0,2^L)$. For example, the hash function mapping
all strings to a constant ($h(s)=z$) has a corresponding compression function $F$ which is defined only
as $F(z,c)=z$ for all characters $c$. It is trivially permuting over a single hash value $z$.)

We have that XOR universality or pairwise independence implies a fixed collision probability of $1/2^L$ between distinct strings. 
This, in turn, implies permuting compression functions by the next Lemma. 

\begin{lemma}\label{lemma:mustbepermuting}An iterated hash family with fixed collision probability ($P(h(s)=h(s'))=\epsilon$ for $s\neq s'$) over strings of length two or more
has \emph{permuting} compression functions. 
\end{lemma}
\begin{proof}
Consider two distinct strings $s,s'$. Consider any iterated hash family with fixed collision probability $\epsilon$. We have 
\begin{eqnarray*}\epsilon & = &  P(h(s\parallel\texttt{c}) = h(s'\parallel\texttt{c}))\\
 & = & P(h(s) = h(s')) +  P(h(s) \neq  h(s')\; \land \; h(s\parallel\texttt{c}) = h(s'\parallel\texttt{c})) \\ 
 & = & \epsilon + P(h(s) \neq  h(s') \; \land \; h(s\parallel\texttt{c}) = h(s'\parallel\texttt{c})).
\end{eqnarray*}
Thus, we have  $h(s) \neq  h(s') \Rightarrow h(s\parallel\texttt{c}) \neq h(s'\parallel\texttt{c})$ which proves the result.
\end{proof}

Consider the consequences of this lemma for $L=1$. Over $\{0,1\}$, there are only 
two permutations: the identity and an exchange of the two values (0 and 1). 
Hence, we have that $F(F(y,\texttt{a}),\texttt{b})=F(F(y,\texttt{b}),\texttt{a})$ if $F$ is permuting. Therefore, the strings \texttt{ab} and \texttt{ba} always collide ($h(\texttt{ab})=h(\texttt{ba})$). 
Thus---in general---XOR universality or pairwise independence over strings longer than $L$~characters is impossible.

However, permuting compression functions have benefits on their own, beside being a consequence of pairwise independence. Consider any fixed permuting compression function $F$ and
any string $s$. Consider any two distinct initial values $H_0$ and $H'_0$. Then the hash value of $s$ computed with $H_0$ must
differ from the hash value computed with $H'_0$ by induction on the number of characters in the string $s$. Thus, we have
the following result. It holds true for strings of arbitrary length.

\begin{lemma}An iterated hash family with permuting compression functions and independently chosen equiprobable initial values 
is uniform.
\end{lemma}

A compression function is \emph{strongly permuting} if it is permuting and if $F(y,c)=F(y,c')$ implies $c=c'$.
Strong permutation means that strings having a Hamming distance of one never collide. Of course, this
precludes pairwise independence. 

\begin{lemma}Given a strongly permuting iterated hash family, two strings differing by exactly one
character never collide.
\end{lemma}

\section{Iterated hash families over variable-length strings}
\label{sec:review}

 We are interested in hashing variable-length strings using iterated hash functions. Let $\Sigma$ be the set of all characters from which the
strings are constructed; the number of distinct characters is $| \Sigma |$. 
We present a range of iterated families (see Tables~\ref{fig:mainuniversality} and~\ref{fig:maincost}). 
Other hash families appear in Appendix~\ref{sec:pophashfunctions}.

\begin{table}[tb]
\caption{\label{fig:mainuniversality} Universality of some iterated families: $n$ is the maximal string length}
\centering\begin{tabular}{cl}\hline
family & universality \\ \hline
\rule{0cm}{1.2em}\textsc{CWPoly} & $n/2^L$-almost XOR universal\\
\rule{0cm}{1.2em}\textsc{Tabulated} & pairwise independent for $n\leq L$ \\ 
\textsc{ShiftTabulated} & pairwise independent on last $L-n+1$~bits \\
\rule{0cm}{1.2em}\multirow{1}{*}{Pearson on unary strings} & $\max_{i<n }d(i)/2^L$-almost universal  \\
\hline
\end{tabular}
\end{table}

\begin{table}[tb]
\caption{\label{fig:maincost} Computational complexity (per character) and memory usage in bits of some iterated families. There
are $|\Sigma |$~characters in the alphabet.}
\centering\begin{tabular}{ccc}\hline
families  & complexity & memory usage \\\hline
\rule{0cm}{1.2em} \textsc{CWPoly} & $O(L \log L 2^{O(\log^*L)})$ & $L$ \\
\rule{0cm}{1.2em}\textsc{Tabulated} and \textsc{ShiftTabulated} & $O(L)$ & $|\Sigma | L$ \\
\rule{0cm}{1.2em}Pearson & $O(L)$ & $\log (2^L! ) \leq L 2^L$ \\\hline
\end{tabular}
\end{table}

\subsection{Carter-Wegman Polynomial Hashing}
\label{sec:cwpoly}
Carter and Wegman~\cite{carter1979uch} defined an almost universal 
family of iterated hash functions (henceforth \textsc{CWPoly}) using polynomials over a finite field $\mathbb{F}_p$
as $h(s)=\sum_i t^{n-i} s_i $ for $t\in \mathbb{F}_p$ randomly chosen and where $s_i$ is the $i^{\mathrm{th}}$~character of the string $s$.
In other words, we use the compression function $F(y,c)= ty+c$.  Characters are interpreted as elements of $\mathbb{F}_p$.
We are especially interested in binary fields where $p=2 ^L$.

Unfortunately, \textsc{CWPoly} with an initial value of zero is such that
 $h(00)=h(0)=0$.
To fix this problem, we need to choose a non-zero initial value~\cite{krovetz2001fast} such as 1. 
Thus, we have $h(00)= t^2$, $h(0)=t$ and $h(223)=t^3+2 t^2+2 t+ 3$.
Therefore, given two strings $s$ and $s'$, $h(s)-h(s')$ is a non-zero polynomial
of degree at most  $\max(|s|,|s'|)$ where $|s|$ and $|s'|$ are the lengths of strings
$s$ and $s'$.
By the Fundamental Theorem of Algebra, such a polynomial has at most $\max(|s|,|s'|)$~solutions.
Thus, the probability of collision between two strings of length at most $n$ is at most $\frac{n}{p}$. 
This probability bound is tight. Indeed, consider the polynomial of degree $n$ over  $\mathbb{F}_p$,  $\tau(t)=\prod_{i=0}^{n-1} (t-i)$. E.g., $\tau(t)=t^3+2t$ for $n=3$ in $\mathbb{F}_3$. It has the $n$~distinct roots 0, 1, \ldots, $n-1$. If $s$ is the character 0 repeated $n$ times, and $s'$ is the string
corresponding to the coefficients of the polynomial $\tau(t)$, we have that $h(s)-h(s')=\tau(t)$, hence $P(h(s)=h(s'))=\frac{n}{2^L}$.
Moreover, Nguyen and Roscoe show that \textsc{CWPoly} has optimal universality given the size of its family~\cite{Ros128}. 
 Sadly, \textsc{CWPoly} is not uniform, but the probability of any hash value $y$ is bounded: $P(h(s)=y)\leq \frac{n}{2^L}$ for any string $s$ of length $n$.

 When the field size is a power of two ($p=2^L$), then  \textsc{CWPoly} is $n/2^L$-almost XOR universal. Indeed, we
 have that $P(h(s) \oplus h(s')=y)$ (for any $y$) is given by the probability that
 $h(s)+h(s')=y$ as polynomials in $\textrm{GF}(2)[x]/p(x)$. Yet $h(s)+h(s')$ is a non-zero polynomial of degree at most  $\max(|s|,|s'|)$ in $t$ and 
 the result follows again by the Fundamental Theorem of Algebra.
 
 Unfortunately, \textsc{CWPoly} cannot be almost strongly universal 
 over variable-length strings even if we
 use a random initial value.
 Indeed, consider the equations $h(\texttt{aa})=\texttt{a}$ and $h(\texttt{a})=0$.
 They can be written explicitly as $H_0 t^2 + \texttt{a} t + \texttt{a} = \texttt{a}$ 
 and $H_0 t + \texttt{a} = 0$ where $H_0$ is the initial value.
 We see that $h(\texttt{a})=0$ implies 
$h(\texttt{aa})=\texttt{a}$. Therefore, if  \textsc{CWPoly}
was $\epsilon$-almost strongly universal for $\epsilon<1$, we would
have that $\epsilon/2^L \geq P(h(\texttt{aa})=\texttt{a} \land h(\texttt{a})=0) = P(h(\texttt{a})=0) = 1/2^L$,
a contradiction. 
It is still possible to modify  \textsc{CWPoly} so that it becomes almost strongly universal.
Unfortunately,  the result is not an iterated family by our definition.
 To get this stronger property, we append to each string a random character
 before hashing.
 We state the general result over $\mathbb{F}_p $, but it obviously
 applies when $p=2^L$.
 
 \begin{lemma}Consider the family \textsc{CWPoly} with 
 the parameter $t$ chosen randomly among the non-zero values of $\mathbb{F}_p $ and an initial value of 1.
 Moreover, we choose a random value $\zeta$ in $\mathbb{F}_p $ and append it to 
 all strings before hashing them:
 \begin{eqnarray*}h(s)=t^{|s|+1}+\sum_{i=1}^{|s|} t^{i} s_i +\zeta\end{eqnarray*}
 If 
we consider strings of length at most $n$, then this modified \textsc{CWPoly} is $(n+1)/(p-1)$-almost strongly universal.
  \end{lemma}
  \begin{proof}With the random parameter $\zeta$,  \textsc{CWPoly} is clearly uniform.
  
  It remains to show that 
  \begin{eqnarray*}P(h(s)=y \land h(s')=y') \leq \frac{n+1}{(p-1) p}\end{eqnarray*}
  for any two distinct strings $s,s'$ and any
  hash values $y,y'$.
We have that $h(s)-h(s')-y-y'$ is a non-zero polynomial of degree at most $n+1$.
  (E.g., if $s = \texttt{ab}$ and $s' = \texttt{cd}$ then
  $  h(s)-h(s')-y-y' = t^2 ( \texttt{a} -\texttt{c}) +  t ( \texttt{b}- \texttt{d} ) -y -y'$.)
  To see why it must be a non-zero polynomial, consider two cases.
  If $s$ and $s'$ have the same length, let $i$ be such that $s_i \neq s'_i$ then $h(s)-h(s')-y-y'$
  as a polynomial over the variable $t$ has $(s_i-s'_i) t^{i+1}$ as its $i+1^{\mathrm{th}}$~term.
  If $s$ and $s'$ have different lengths, assume without loss of generality that $s$ is
  longer, then the $|s|+1^{\mathrm{th}}$~term of the polynomial is $t^{|s|+1}$ and therefore
  the polynomial has degree $|s|+1$ and is non-zero.
  Hence, the polynomial  has at most  $n+1$~roots.
  Moreover, we have that $h(s)-h(s')-y-y'$ is independent from $\zeta$.
  Given any $t$ such that $h(s)-h(s')-y-y'=0$ is satisfied, there is only one value $\zeta$ (dependent on $t$) such that $h(s)=y$.
  Thus there are at most $n+1$~pairs of values $t,\zeta$ such that $t\neq 0$,
$h(s)-h(s')-y-y'$ and $h(s)=y$ are satisfied. Therefore, the probability that
$h(s)=y$ and $h(s')=y'$ are both true is bounded by $\frac{n+1}{(p-1)p}$ because there are
$p-1$~possible non-zero values for $t$ and $p$~possible values for $\zeta$.
  \end{proof}

When $L$~bits fit in a processor register, the running time of the  compression function $F$ may be considered independent of $L$.
More formally, however, the multiplication between two $L$-bit integers required by the compression function is in $O(L \log L 2^{O(\log^*L)})$~\cite{1250800}.

\subsection{Iterated string hashing by tabulation}
\label{sec:iteratedbytabulation}
Hashing by tabulation~\cite{thorup2004tabulation,cohenhash,carter1979uch,lemi:one-pass-journal}
has good universality, at the expense of the memory usage. We adapt this strategy to iterated hashing of variable-length strings.

Consider $\Gamma$, a randomly chosen function from characters in $\Sigma$ to  $L$-bit hash values. 
There are $2^{L | \Sigma | }$~such functions.
We consider integers as elements of $\mathbb{F}_{2^L}$, that is, as polynomials
with binary coefficients
in $\textrm{GF}(2)[x]/p(x)$ where $p(x)$ is an arbitrarily chosen irreducible
polynomial of degree $L$.
Consider the family of iterated hash functions with compression functions
of the form $F(y,c)=x y + \Gamma( c)$ and an initial value chosen randomly (henceforth
\textsc{Tabulated}). The compression function is permuting.
\textsc{Tabulated} is pairwise independent.

\begin{proposition}\label{prop:tabulated}\textsc{Tabulated} is pairwise independent for
strings no longer than $L$~characters.
\end{proposition}
\begin{proof}
Consider two distinct strings $s, s'$ no longer than $L$~characters. 
We want to show that $P(h(s)=y \; \land \; h(s')=y')= 1/2^{2L}$ for any hash values $y,y'$.

Consider the equation $h(s)=y$ or \begin{eqnarray*}H_0 x^{|s|} + \sum_{i=1}^{|s|} x^{|s|-i} \Gamma(s_i) = y\end{eqnarray*} where $H_0$ is the initial value. We
can solve for the initial value as a function of $y$ and $s$:  
\begin{eqnarray}H_0 = x^{-|s|} (y -  \sum_{i=1}^{|s|} x^{|s|-i} \Gamma(s_i)).\label{eqn:h0}\end{eqnarray}

We  solve for $\Gamma(s_j)$ for one value of $j$, in terms that do not
depend on $H_0$. This will allow us to conclude the proof. We consider two cases:
\begin{itemize}
\item Suppose that the strings have the same length ($|s|=|s'|$).
From  $h(s) = y$ and  $h(s') = y'$, we get that  $h(s)- h(s') = y-y'$.
We have that the equation $h(s)- h(s') = y-y'$ is independent from $H_0$ because
the terms $H_0 x^{|s|}$ and $H_0 x^{|s'|}$ cancel out.
Because the strings are distinct, there must be an index $j\in\{1,2,\dots, |s|\}$
such that $s_j \neq s'_j$.
Let the character $s_j$ occur at indexes $r_1, r_2, \dots, r_k$ in string
$s$ (by definition $j\in\{ r_1, r_2, \dots, r_k\}$)
and at indexes $r'_1, r'_2, \dots, r'_l$ in string $s'$.
If we let $q= \sum_m x^{|s|-r_m} - \sum_m x^{|s|-r'_m}$,
the equation
$h(s)- h(s') = y-y'$ can be written as $q \Gamma(s_j) = \lambda$ for
some value $\lambda\in \mathbb{F}_{2^L}$ independent from $\Gamma(s_j)$ and $H_0$.
Because $j$ is in $\{ r_1, r_2, \dots, r_k\}$ but not in  $\{ r'_1, r'_2, \dots, r'_l\}$,
and because all $|s|-r_m$'s and all $|s|-r'_m$'s are less than $L$,
we have that $q\neq 0$.
 \item Suppose that the strings have different lengths. 
Without loss of generality,
assume that $|s|>|s'|$.
From  $h(s) = y$ and  $h(s') = y'$, we get that  $(h(s)- y) - x^{|s|-|s'|}(h(s')-y') =0$.
The equation $(h(s)- y) - x^{|s|-|s'|}(h(s')-y') =0$ is independent from $H_0$
because terms $H_0 x^{|s|}$ and $H_0 x^{|s'|} \times x^{|s|-|s'|}$ cancel out.
Consider the character $s_{|s|}$ and seek all indexes where it appears:
write these indexes 
 $r_1, r_2, \dots, r_k$ for string
$s$ (by definition $|s|\in\{ r_1, r_2, \dots, r_k\}$)
and $r'_1, r'_2, \dots, r'_l$ for string $s'$.
We have that  $q= \sum_m x^{|s|-r_m} - \sum_m x^{|s|-r'_m  }$
is non-zero because $0$ is in $\{ |s|-r_1, |s|-r_2, \dots, |s|-r_k\}$ but
not in $\{|s|-r'_1, |s|-r'_2, \dots, |s|-r'_l \}$ since $r'_m\leq |s'|<|s|$ for all $m$'s, and because the $|s|-r_m$'s and the $|s|-r'_m$'s are less
than $L$. For the rest of the proof, we set $j=1$.
\end{itemize}
Hence we can solve for $\Gamma(s_j)$ as $\Gamma(s_j)=q^{-1} \lambda$
whether the two strings have the same length or not.  
Equation~\ref{eqn:h0} gives $H_0$ as a function of
$ \Gamma(s_i)$ for $i =\{1,2, \ldots, |s|\}$. So, our formula
for $H_0$ depends on $ \Gamma(s_j)$, but we can substitute $\Gamma(s_j)=q^{-1} \lambda$
in this formula (Equation~\ref{eqn:h0}) to get an expression for $H_0$ which does
not depend on $ \Gamma(s_j)$.
Thus, from the equations $h(s)=y$ and $ h(s')=y'$, we get one and only one value for $H_0$ and
$\Gamma(s_i)$ as a function of the other tabulated values and of $y$ and $y'$.
Both values are chosen at random among $2^L$~values and thus
the result is shown.
\end{proof}

For binary strings, \textsc{Tabulated} has nearly optimal memory-universality trade-off.
Indeed, there are $2^{L+1}-2$~binary strings of length at most $L$. Hence, any pairwise independent
family over such binary strings must contain at least $1+(2^{L+1}-2)(2^L-1)$~hash functions~\cite{188765}.
Therefore, its cardinality must be in $\Omega(2^{2L})$. Meanwhile, \textsc{Tabulated} 
has $2^{2L}$~hash functions over binary strings.

\subsection{Iterated string hashing by shifted tabulation}

While  \textsc{Tabulated} is pairwise independent, it requires operations in finite fields of cardinality $2^L$. 
Thankfully some microprocessors have instructions for computations in such finite fields~\cite{intelcarryless}. 
Yet it can be
expensive on some computers, even with such instructions. Fortunately, pairwise independence is possible without finite fields~\cite{dietzfelbinger1996universal}.

The barrel  or circular shift  is the invertible operation by which all bits all shifted, except for the last ones, which are brought back at the beginning.  For $L$-bit values, the barrel left shift by one can be written as
$y\circlearrowright 1 = (y\ll 1) \oplus  (y\gg L-1)$  where $\ll$ and $\gg$ are the left and right shifts.
E.g., $11001$ becomes $10011$.
Barrel shifting can be implemented efficiently in hardware~\cite{252605}. The popular x86 and ARM instruction sets offer the \texttt{ror}  instruction for this purpose.

  Consider the hash
family  (henceforth \textsc{ShiftTabulated}) with compression functions of the form 
$F(y,c)= (y\circlearrowright  1) \oplus   \Gamma( c)$ where 
$\Gamma$ is a randomly chosen function from  characters to $L$-bit hash values. 
(Whether we choose the barrel left or right shift is arbitrary.)
We choose the initial value randomly.
Because the
 compression function is permuting, \textsc{ShiftTabulated} is uniform.

The  \textsc{ShiftTabulated} compression functions can be computed efficiently: one value to look-up, one barrel shift and one  bitwise XOR\@.
Moreover, \textsc{ShiftTabulated} can be described using the same compression function as \textsc{Tabulated}---$F(y,c)=x y + \Gamma( c)$---in
$\textrm{GF}(2)[x]/(x^L+1)$. (The polynomial $x^L+1$ fails to be irreducible and thus, $\textrm{GF}(2)[x]/(x^L+1)$ is merely 
a ring.)

The proof of the pairwise independence of \textsc{Tabulated} (see Proposition~\ref{prop:tabulated}) relies on the fact that
any non-zero element of the field $\textrm{GF}(2)[x]/p(x)$ invertible, which includes any non-zero polynomial of degree at most $L-1$ from $\textrm{GF}(2)[x]$. 
In turn, this means that given any polynomial $q$ of degree at most $n-1<L$, and any element $r$ of the field, we have
\begin{eqnarray*}P(q \Gamma(s_i) =  r) = 2^{-L}\end{eqnarray*}
whenever $\Gamma(s_i)$ is picked at random in the field, because the equation is only true when $ \Gamma(s_i) =  q^{-1 }r$.
The same is \emph{almost} true in $\textrm{GF}(2)[x]/(x^L+1)$.
We write that two values are equal modulo the first $n-1$~bits if we ignore the first $n-1$~bits in the comparison.
By Corollary~1 from Lemire and Kaser (2010)~\cite{lemi:one-pass-journal},
we have that \begin{eqnarray*}P(q \Gamma(s_i) =  r \bmod{\textrm{first $n-1$~bits}}) = 2^{-L+n-1}.\end{eqnarray*}  Hence, by a proof
similar to Proposition~\ref{prop:tabulated}, we have the following result.

\begin{lemma}\label{prop:shiftedtabulated}\textsc{ShiftTabulated} is pairwise independent on the last $L-n+1$~bits for
strings no longer than $n$~characters. 
\end{lemma}

\subsection{Pearson hashing}
\label{sec:pearson}

We define Pearson hashing by the family of compression functions $F(y,c)=A_{y \oplus c}$ where $A$ is an array containing
a permutation of the values in $\{0,1,\dots,2^L-1\}$~\cite{78978}.
These compression functions are strongly permuting: $A_{y \oplus c}=A_{y' \oplus c}$ implies $y=y'$, $A_{y \oplus c}=A_{y \oplus c'}$ implies $c=c'$.
We pick the initial value uniformly at random. Thus, Pearson is uniform. To our knowledge,
the exact universality of Pearson remains unknown.
(We know that it can never be strongly universal because its compression function
is strongly permuting.) 


For  $L=2$, Pearson is 5/6-almost universal for strings no longer than four. 
That is, it is universal
for strings of length $2^L$ unlike \textsc{CWPoly}. Brute-force numerical investigations for large values
of $L$ is difficult.

To simplify the analysis, we
focus on unary strings: strings made of a single character (such as \texttt{aaaa}).
The following result is an upper bound on the universality of Pearson over general strings.

\begin{proposition}\label{prop:pearson}Pearson is $\epsilon$-almost universal over unary strings of length at most $n$ for $\epsilon = \max_{i<n }d(i)/2^L$ where 
$d(i)$ is the divisor function---the number of positive integers dividing $i$.
\end{proposition}
\begin{proof}
Consider strings made of the character $\texttt{a}$.
Let $\pi$ be the permutation $A_{\cdot  \oplus \texttt{a}}$. 
Fix the initial value $H_0$.
A collision between two unary strings of lengths $k,k'\leq n$ is equivalent to the equation $\pi^{k} H_0 = \pi^{k'} H_0 \Rightarrow \pi^l H_0 = H_0$ for $|k-k'|=l< n$.
Consider any solution $\varpi$ of this equation ($\varpi^l H_0 = H_0$), then let $\sigma$ be the smallest integer such that $\varpi^{\sigma} H_0 = H_0$. We bound the number of solutions using the fact that $\sigma$ must divide $l$. 

Given $\sigma$, there are $(2^L-1)!$~solutions to the equation $\pi^{\sigma} H_0 = H_0$ subject to $\pi^{i} H_0 \neq H_0$ for $i<\sigma$. Indeed, we have that $\pi H_0$ can be any value, except $H_0$---thus we have $2^L-1$~possibilities. We have that
$\pi^2 H_0$ can be any value expect $H_0$ and $\pi H_0$, hence we have $2^L-2$~possibilities. And so on, up to 
$\pi^{\sigma} H_0$ which is predetermined. 
At that point, we have enumerated $(2^L-1) (2^L-2) \cdots (2^L-\sigma+1)$~possibilities. Each one of these possibilities define how the values $H_0, \pi H_0, \dots, \pi^{\sigma-1} H_0$ are permuted.
 The other $2^L-\sigma$~values can be permuted to any available value, generating $(2^L - \sigma)!$~possibilities for $(2^L-1)!$~possibilities.

Thus there is a total of $d(l) (2^L-1)!$~solutions and $(2^L)!$~different permutations: 
the ratio is \begin{eqnarray*}\frac{d(l) (2^L-1)!}{(2^L)!} = \frac{d(l)}{2^L}.\end{eqnarray*} This is true for any initial value $H_0$. 
The string length difference $l$ ranges between 1 and $n-1$: we must keep the maximum value $\max_{i<n }d(i)/2^L$.
\end{proof}

\begin{figure}\centering
\includegraphics[height=0.6\textwidth,angle=270]{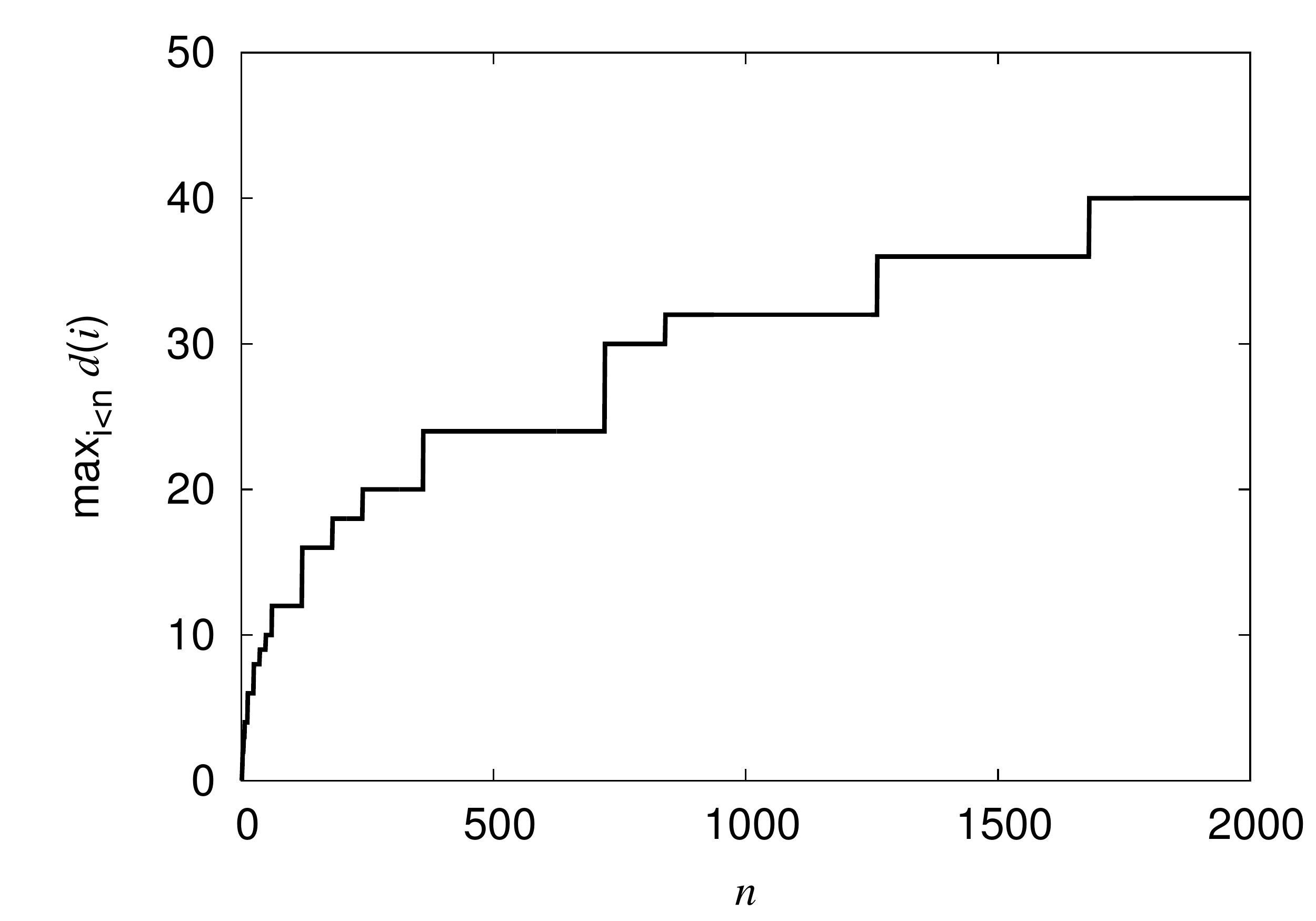}
\caption{\label{fig:divisor}Plot of  $\max_{i<n }d(i)$ where $d$ is the divisor function}
\end{figure}

The  function $\max_{i<n }d(i)$ grows slowly  (see Fig.~\ref{fig:divisor}). 
Formally, we have that   $\max_{i<n }d(i) \in o(i^\epsilon)$ for all $\epsilon>0$. 
For $\max_{i<n }d(i)$ to be equal to $2^L$---so that Pearson is no longer almost universal over unary strings---we need
$n$ to be larger than the  least common multiple of the integers from 1 to $2^L$. Thus, Pearson
can be almost universal over unary strings much longer than $2^L$~characters.

To prove that  iterated hashing for non-unary strings longer than $2^L$~characters
is possible,  
we 
consider the following variation on Pearson hashing (henceforth \textsc{Generalized Pearson}):
we pick compression functions of the form $F(y,c)=A_{y \oplus c}$ where $A$ is a random array
containing values in $\{0,1,\dots,2^L-1\}$ (not necessarily a permutation).
For $L=1$, 
we have that $h(00)=h(11)$ with probability one.
However, for $L=2$, \textsc{Generalized Pearson} is almost universal for
strings longer than $2^L$ (see Table~\ref{table:gpearson}). We computed 
these probabilities by enumerating all possible compression functions
and all possible pairs of strings with $L$-bit characters.
\begin{table}[tb]
\caption{\label{table:gpearson}Numerically-derived upper bound on the collision probability between strings of length at most $n$ under \textsc{Generalized Pearson} ($L=2$, $2^L=4$)}\centering
\centering\begin{tabular}{ccc}\hline
 $n$ & collision probability \\ \hline
2 & 0.53 \\
3 & 0.72\\
$4=2^L$ & 0.84\\\hdashline[1pt/5pt] 
5 & 0.88\\
6 & 0.89\\
7 & 0.95\\ 
8 & $\geq 0.97$ \\ 
9 & $\geq 0.98$ \\ 
10 & $\geq 0.99$ \\ 
11 & 1.00 \\ 
\hline
\end{tabular}
\end{table}

\section{Bounding the universality of iterated hashing by the maximal family size}

\label{sec:nguyenroscoe}

Given only the number  of hashable items, the number of hash values
and the number of hash functions, we can bound the universality.
To be $\epsilon$-almost universal, a family must have enough hash functions.

There is a limited number of iterated hash functions.
Compression functions have
$2^{L} |\Sigma | $~possible inputs. For each input,
there are $2^L$~possible hash values. Thus, there
are no more than $2^{L 2^L |\Sigma |}$~compression
functions. Moreover, there are no more than $2^L$~initial values $H_0$. Thus the number of
iterated hash functions $| \mathcal{H} |$ is bounded by $2^{L (2^L |\Sigma |+1)}$:
\begin{eqnarray}\label{eqn:weakbound}
| \mathcal{H} |\leq 2^{L (2^L |\Sigma |+1)}.
\end{eqnarray}
There are  
$|\Sigma |^n+|\Sigma |^{n-1}+\cdots +|\Sigma |$~possible
non-empty strings. 

Nguyen and Roscoe derived a bound which is particularly suited for values of $\epsilon$ larger than $1/2^L$~\cite{Ros128}.
Let $2^K$ be the number
of hashable items. Pick any hash
function. By the pigeon-hole principle, there must at least $2^{K}/2^L$~hashable items
colliding to the same value. Apply any other hash function to these colliding items,
there must be $2^{K}/2^{2 L}$~items colliding on these two hash functions. By
repeating this argument, Nguyen and Roscoe show that $\epsilon$-almost universal hashing requires  $\lceil K/L - 1\rceil / \epsilon$~hash functions~\cite{Ros128}.

\begin{corollary}[Nguyen and Roscoe]\label{corollary:NguyenRoscoe} Given at least $2^K$~hashable items, then $\epsilon$-almost universal hashing for $\epsilon>0$ requires at
least $\lceil K/L - 1\rceil / \epsilon$~hash functions.
\end{corollary}
\begin{proof}
We have $X=2^r$~hash functions.
Theorem~1 from Nguyen and  Roscoe~\cite{Ros128} states that when $K$ is a multiple of $L$ then $r\geq \log (\epsilon^{-1} (\frac{K}{L}-1))$ and 
$r\geq \log (\epsilon^{-1} \lfloor \frac{K}{L}\rfloor)$ otherwise. 
Their result may be rewritten as $r\geq \log (\epsilon^{-1} (\lceil \frac{K}{L}- 1 \rceil ))$, irrespective of the value of $K$. Thus, we have
$X=2^r \geq \lceil \frac{K}{L} - 1\rceil /    \epsilon$. It proves the result.
\end{proof}

By combining  this last corollary with our bound on the number of iterated hash functions
(see Equation~\ref{eqn:weakbound}), we get the following bound on the universality of
iterated hashing.
 
\begin{lemma}\label{lemma:countingbased}At best, iterated hashing might be 
$\epsilon$-almost universal over the strings of length at most
$L(\epsilon 2^{L (2^{L+1} +1)} +1)$ for some $\epsilon<1$.
\end{lemma}
\begin{proof}
According to Corollary~\ref{corollary:NguyenRoscoe}, we have that
 $|\mathcal{H}|\geq \lceil K/L - 1\rceil / \epsilon$.
 Solving for $K$ in this expression, we get $K\leq L (\epsilon | \mathcal{H} | +1)$. 
 Meanwhile, for string of length at most $n$ over $|\Sigma|$~distinct characters, 
 we have $|\Sigma |^n \leq 2^K$ or $ n \log |\Sigma | \leq K$.
 Hence, by combining these two inequalities, we have 
 $n \log |\Sigma |  \leq  K \leq L (\epsilon | \mathcal{H} | +1)$
 or just 
\begin{eqnarray*}n \log |\Sigma |  \leq L (\epsilon | \mathcal{H} | +1).
\end{eqnarray*}
 
  Moreover, we can bound the size of an iterated family 
 as $|\mathcal{H}|\leq 2^{L (2^L |\Sigma |+1)}$ (see Equation~\ref{eqn:weakbound}).
Hence, by substitution,  we have 
\begin{eqnarray*}
 n \log |\Sigma | & \leq  & L (\epsilon | \mathcal{H} | +1)\\
   & \leq  & L (\epsilon 2^{L (2^L |\Sigma |+1)} +1).
\end{eqnarray*}
Finally, we can solve for $n$ in this inequality.
Thus---at best---iterated hashing might be
almost universal over strings of length at most
$L(\epsilon 2^{L (2^L |\Sigma |+1)} +1)/\log |\Sigma |$
where $|\Sigma |>1$.
This bound grows exponentially with $|\Sigma |$
which is misleading because universality over a large alphabet ($|\Sigma |$ large)
implies universality over a smaller alphabet. Indeed, it is always possible to
restrict the application of a universal family to strings using few characters, and this
restriction may only increase the universality.
Thus, it is preferable to set $|\Sigma |=2$.
(For $| \Sigma |=1$, we
get a weaker bound: 
 we have that $2^K \geq n$ and $\mathcal{H} \leq 2^{L2^L}$ so
that the bound becomes $n\leq 2^{L \epsilon 2^{L2^L}+1}$.)
\end{proof}

For bounding universality, that is $1/2^L$-almost universality, it is preferable
to use  bound provided by Stinson~\cite{188765} to get the following result. 


\begin{lemma}\label{lemma:stinson}At best, iterated hashing might be 
 universal  over the strings of length at most
$2 L+L 2^{L+1}$.
If the family is strongly universal,
then it is limited to strings of at most
$L + 2 \log 2^L! - \log(2^L-1)-1$~characters.
\end{lemma}
\begin{proof}
First, we consider universal hashing.
By Equation~\ref{eqn:weakbound}, we have that $2^{L(2^L | \Sigma| +1)} \geq | \mathcal{H} |$.
Stinson~\cite{188765} proved that the size of universal families must be at least
as large as the number of hash values divided by the number of elements, thus we have 
$| \mathcal{H} | \geq | \Sigma |^n / 2^L$. By combining these two inequalities,
we get 
\begin{eqnarray*}
2^{L(2^L | \Sigma| +1)} \geq | \mathcal{H} | \geq | \Sigma |^n / 2^L
\end{eqnarray*}
or 
\begin{eqnarray*}
2^{L(2^L | \Sigma| +1)} \geq  | \Sigma |^n / 2^L.
\end{eqnarray*}
Taking the logarithm on both sides, we get 
\begin{eqnarray*}
 n \leq \frac{L(2^L | \Sigma| +2) }{\log | \Sigma |}.
\end{eqnarray*}
The right-hand-side of this last inequality grows with $|\Sigma|$,
but a family universal over a large alphabet must be universal over a smaller
alphabet as well. Thus we set $|\Sigma| =2$. This proves the first part of the lemma.

Consider strongly universal hashing.
Recall that Stinson~\cite{188765} proved that strongly universal families 
must have cardinality at least $1+a(b-1)$ where $a$ is the number of strings
and $b$ is the number of hash values. Hence, we have that 
$| \mathcal{H} | \geq 1 + | \Sigma |^n ( 2^L - 1)$.
As a consequence of Lemma~\ref{lemma:mustbepermuting}, strongly universal 
hash families must have permuting compression functions.
There are $(2^L!)^{ | \Sigma |}$~such functions, and $2^L$~possible initial
values for a total of at most $2^L \times 2^L!^{ | \Sigma |}$~hash functions.
Hence, we have $2^L \times 2^L!^{ | \Sigma |} \geq |\mathcal{H}|$.
By combining these inequalities, we get 
\begin{eqnarray*}
2^L \times 2^L!^{ | \Sigma |}   \geq 1 + | \Sigma |^n ( 2^L - 1).
\end{eqnarray*}
We can drop the constant term 1 from the right to get
\begin{eqnarray*}
 | \Sigma |^n < \frac{2^L \times 2^L!^{| \Sigma | }}{ 2^L - 1}.
\end{eqnarray*}
As before, we can set $| \Sigma |=2$ to 
get
\begin{eqnarray*}
 n < L + 2 \log 2^L! - \log(2^L-1).
 \end{eqnarray*}
 This concludes the proof.
\end{proof}

\section{Limitations of iterated hashing over long strings}
\label{sec:limitations}

To characterize the
limitations of iterated hashing---irrespective of the family size, 
we want to compute a bound on the string length given a desired bound $\epsilon$ on the collision probability.

Let $s_{r,\texttt{a}}$ be the unary string made of the character \texttt{a} repeated $r$~times.
For example, we have $s_{3,\texttt{a}}=\texttt{a}\texttt{a}\texttt{a}$. 
 Because
we have at most $2^L$ distinct hash values, we have that
$h(s_{2^L+1,\texttt{a}})$ must be equal to $h(s_{r,\texttt{a}})$ for some $r\in \{1,\ldots, 2^L\}$.
Hence, we have the following lemma.

\begin{lemma}\label{lemma:tech11}For any iterated hash function $h$ and any character \texttt{a}, the values $h(s_{r,\texttt{a}})$  are cyclic over $r\geq 1$ with a period $T\in \{1,2,\ldots,2^L\}$ except maybe for the first $2^L-T$~hash values.
\end{lemma}
\begin{proof}
In the $2^L+1$~hash values $h(s_{r,\texttt{a}})$ for $r\in \{1,2,\ldots,2^L,2^L+1\}$, one value must be repeated
because there are at most $2^L$~distinct hash values. 
Write $h(s_{r_1,\texttt{a}})=h(s_{r_2,\texttt{a}})$, then by Proposition~\ref{prop:equiv},
$h(s_{r_1+i,\texttt{a}})=h(s_{r_2+i,\texttt{a}})$ for any non-zero integer $i$.
Without loss of generality, assume $r_2>r_1$.
This proves that
$h(s_{r_1+x,\texttt{a}})$ is cyclic in $x$ with period at most $T=| r_1-r_2| $. We see that $1\leq T \leq 2^L$.
Only the $h(s_{i,\texttt{a}})$ for $i=1,2,\ldots, r_1-1$ are excluded from our analysis. This concludes the proof.
\end{proof}

Let $LCM_{k}\equiv LCM(\{1,2,\ldots,k\})$ be the least common multiple of the integers from 1 to $k$, inclusively. For example, we have $LCM_{2}=2,LCM_{4}=12, LCM_{8}=840$. 
By definition, for any $T\in \{1,2,\ldots,k\}$, we have that $T$ divides $LCM_{k}$.
Thus, the strings
$s_{2^L,\texttt{a}}$ and $s_{2^L+LCM_{2^L},\texttt{a}}$ collide
with probability one under iterated hashing by Lemma~\ref{lemma:tech11}.
Using a generalized argument, we have the following proposition.

\begin{proposition}\label{prop:almostlimit}We have the following results  
concerning iterated hashing over variable-length strings:
\begin{itemize}
\item Almost universality  over
strings of length up to $2^L+LCM_{2^L}$ is impossible;
\item Universality over strings of length up to $2^L+2$ is impossible;
\item For $1/2^L<\epsilon<1$ such that $1/\epsilon$ is not an integer, $\epsilon$-almost universality over strings of  length at most
$2^L+ LCM_{2^L+1-\lfloor 1/\epsilon\rfloor}$ is impossible.
\end{itemize}
\end{proposition}
\begin{proof} Since the values $h(s_{r+2^L-1,\texttt{a}})$ must have period $T \in \{ 1, \dots, 2^L\}$ as functions of $r$, and $T$ must divide $LCM_{2^L}$, we have that 
$h(s_{2^L,\texttt{a}})= h(s_{2^L+LCM_{2^L},\texttt{a}})$ for all iterated hash functions $h$ (see Lemma~\ref{lemma:tech11}).
This proves the first result.

We prove the last result. Suppose that hashing is $\epsilon$-almost universal. Then the probability that $h(s_{r+2^L-1,\texttt{a}})$ is 
cyclic in $r$ with period $T$ is bounded by $\epsilon$: $P(\textrm{period}(h)=T)\leq \epsilon$.
Thus, we have $P(\textrm{period}(h)\in \{ 2^L-j+1,2^L-j+2, \dots, 2^L \})\leq j \epsilon$ for any integer $j$ between 1 and $2^L$.
Because $P(\textrm{period}(h) \in \{1,2,\dots,2^L\})=1$, we have
$P(\textrm{period}(h)\leq  2^L-j)\geq 1- j \epsilon$.
Setting $j=\lfloor 1/\epsilon \rfloor - 1 $, we have
that 
$P(\textrm{period}(h)\leq 2^L-j)\geq 1- (\lfloor 1/\epsilon \rfloor - 1) \epsilon =\epsilon - \lfloor 1/\epsilon \rfloor \epsilon >  \epsilon$.
Thus, the probability
$P(h(s_{2^L,\texttt{a}})= h(s_{LCM_{2^L+1-\lfloor 1/\epsilon\rfloor  }+2^L,\texttt{a}}))>\epsilon$
which concludes the proof of the last item. 

The second result follows because universality implies
$1/2^L+\delta$-almost universality for all $\delta>0$. We can find $\delta$ sufficiently small,
such that $1/\epsilon$  is not an integer, and such that $\lfloor 1/\epsilon\rfloor= 2^L-1$. Thus,
we have that universality over strings of length at most $2^L+ LCM_{2}= 2^L +2$ is impossible. This concludes the proof.
\end{proof}

We have that $LCM_{2^L}$ divides $2^L!$ so $LCM_{2^L}\leq 2^L!$; moreover, by a standard identity $2^L!<(2^{L})^{2^L}=2^{L2^L}$.
Hence, the bound on almost universality from this last proposition is preferable to the cardinality-based  bound (see Lemma~\ref{lemma:countingbased}).
Similarly, we compare the bounds on universality in Table~\ref{table:compbounds}: the new bound of $2^L+1$~characters is much smaller.

\begin{table}
\caption{\label{table:compbounds}Comparison of the bounds on universality from Lemma~\ref{lemma:stinson} and Proposition~\ref{prop:almostlimit} }
\centering
\begin{tabular}{c|rr|r}\hline
\multirow{2}{*}{$L$} & \multicolumn{2}{|c|}{Lemma~\ref{lemma:stinson}} & \multicolumn{1}{c}{Proposition~\ref{prop:almostlimit}} \\
 & \multicolumn{1}{|c}{universality}  & \multicolumn{1}{c|}{strong univer.} & \multicolumn{1}{c}{universality} \\\hline
2 & 20 &  8 & 5\\
4 & 136 &87 &17\\
8 & 4\,112 & 3\,366 & 257\\
16 & 2\,097\,184 & 1\,908\,072 & 65\,537\\\hline
\end{tabular}
\end{table}

At least for unary strings, the next lemma shows that the almost universality bound of Proposition~\ref{prop:almostlimit}
is tight (up to one character).
 
\begin{figure}[tb]
\centering
\includegraphics[height=0.6\textwidth,angle=270]{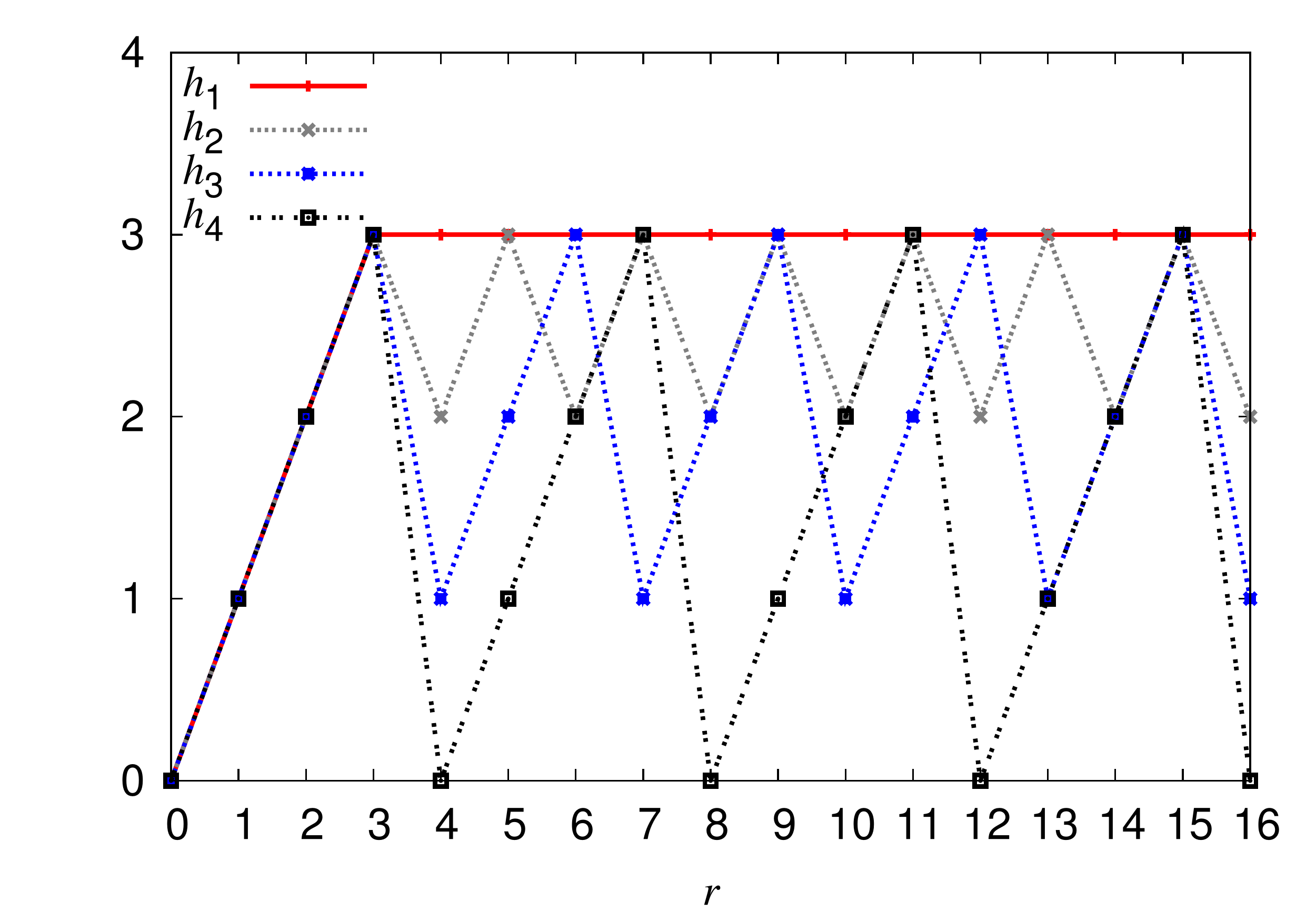}
\caption{\label{fig:cyclicounter}Functions $r\rightarrow h_T(s_{r,\texttt{a}})$ for $L=2$ and $T=1,2,3,4$}
\end{figure}

\begin{lemma}\label{lemma:difflength}
There exists an almost universal iterated family 
over unary strings of length at most  $2^L+LCM_{2^L}-2$.
\end{lemma}
\begin{proof}
For $T\in \{1,2,\ldots,2^L\}$, define 
\begin{eqnarray*}
h_T(s_{r,\texttt{a}})= \begin{cases}
r & \text{if $0\leq r< 2^L$}\\
 2^{L}-T - (r-2^L \bmod{T}) & \text{otherwise}
\end{cases}
\end{eqnarray*}
Effectively, $h_T$ goes from 0 to $2^L-1$ for strings of length $0$ to $2^L-1$, and then
it becomes cyclic with period $T$ (see Fig.~\ref{fig:cyclicounter}).
This family is iterated. 

We want to show that there is no pair of strings $s_{r,\texttt{a}}, s_{r',\texttt{a}}$ for $r,r'\leq LCM_{2^L}+2^L-1$ such that $h_T( s_{r,\texttt{a}} ) = h_T( s_{r',\texttt{a}} )$ for all $T\in \{1,2,\ldots,2^L\}$.
Suppose that it is false. It cannot happen if $r,r'< 2^L$ since $h_T( s_{r,\texttt{a}} ) = h_T( s_{r',\texttt{a}} )$ would imply $r=r'$. Suppose that
$h_T( s_{r,\texttt{a}} ) = h_T( s_{r',\texttt{a}} )$ for all $T\in \{1,2,\ldots,2^L\}$, and for some $2^L\leq r\leq LCM_{2^L}+2^L-2$ and some  $r'< 2^L$. 
Then $h_T( s_{r',\texttt{a}} )=r'-1$.
This would imply that $h_T( s_{r,\texttt{a}} )$ is independent of $T$ which is not possible for $2^L<r<LCM_{2^L}+2^L$ because $h_T( s_{r,\texttt{a}} )$ is cyclic with period $T$.
Similarly, for $2^L\leq r,r'<LCM_{2^L}+2^L-1$,  
the equality $h_T( s_{r,\texttt{a}} ) = h_T( s_{r',\texttt{a}} )$
is possible only when  $T\in \{1,2,\ldots,2^L\}$ divides $|r-r'|$.
But since $|r-r'|<LCM_{2^L}$, this is not possible for all $T\leq 2^L$.

The result is shown.
\end{proof}

Finally, we show that  the  universality bound of  Proposition~\ref{prop:almostlimit} is tight for unary strings (up to two characters). 
To prove the result, we build a  perfect (collision-free) hash function 
over unary strings of length at most $2^L$. 
Consider strings made of the character $\texttt{a}$.
Let $\pi$ be a cycle of length $2^L$ over the integers
$\{1,2,\ldots,2^L\}$. For example, let $\pi$ be
the permutation that takes $k$ to $k+1$ for $k<2^L$ and 
$2^L$ to $1$.
We choose the hash function $h(s_{r,\texttt{a}})= \pi^r 1$. 
A collision between any two strings of
length at most $2^L$ implies that  
$ \pi^l H_0 = H_0$ for some $l< 2^L$ which is impossible because $\pi$ is a cycle of length $2^L$. Thus,
no two unary strings of length at most $2^L$ may collide under this hash function.

\section{Conclusion}

We have shown that iterated hashing can be pairwise independent over short strings.
Moreover, iterated hashing can be almost universal over strings
longer than the number of hash values.
Motivated by this result,
we have derived  bounds on the universality of iterated hashing. 
We can construct large iterated hashing
families: this might suggest that a very high degree of universality
is possible. Alas we have shown that this expectation would be misguided

We have identified two open problems which we find interesting. On the one hand, 
we lack a bound on the universality of iterated hashing given the size of the family.
Our bounds assume arbitrarily large families.
On the other hand, we are still missing provably optimal iterated families.
For example, is it possible to construct a family which
pairwise independent for strings longer than $L$ for some values of $L>1$?
Future work might consider the specific limitations of other hashing strategies~\cite{188765,Sarkar2009}.

\section*{Acknowledgements}

This work is supported by NSERC grant 261437. 

\bibliographystyle{elsart-num-sort}
\bibliography{lemur,zobelhash} 

\appendix

\section{Other popular iterated hash functions}

\label{sec:pophashfunctions}

For completeness, we review some popular iterated hash families and functions.
We show that many of these families have permuting compression functions.

\subsection{Hashing by random irreducible polynomials}
\label{app:irreduciblepolynomial}
Instead of using a fixed irreducible polynomial, as in \textsc{CWPoly}, we can
pick the irreducible polynomial at random~\cite{krawczyk1994lfsr,shoup1996fast} (henceforth \textsc{Division}). 
Considering $L$-bit characters as polynomials having binary coefficients ($\textrm{GF}(2)[x]$),
we use the compression function $F(y,c)=y x^L + c $. As in \textsc{CWPoly}, we specify 
an initial value of 1. Thus, given a string $s$, the hash value is
 $x^{nL} + s_1 x^{(n-1)L} + \dots + s_n \bmod p(x)$. Consider two strings of length at most $n$. 
The equation $h(s)=h(s')$
is true in $\textrm{GF}(2)[x]/p(x)$ only if the non-zero polynomial of degree at most $nL$ formed by $h(s)-h(s')$ is divisible by
$p(x)$. The polynomials of degree $nL$ have at most $n$~irreducible factors of degree $L$.
Meanwhile, there are at least $(2^{L}-2^{L/2+1})/L$~irreducible
polynomials $p(x)$ of degree $L$~\cite{Piret86}. Thus, the probability of a collision is
no larger than $\frac{nL}{2^{L}-2^{L/2+1}}$.

 We can extend this analysis to show almost XOR universality with the same bound ($\frac{nL}{2^{L}-2^{L/2+1}}$). Pick any value $y$, then the probability 
 $P(h(s) \oplus h(s') = y)$ is given by the probability that $h(s)+h(s')-y=0$ in $\textrm{GF}(2)[x]/p(x)$.
 The polynomial  $h(s)+h(s')-y$ in $\textrm{GF}(2)[x]$ has degree at most $nL$ but at least $L$, and the result follows.

The compression function of \textsc{Division} is (strongly) permuting: $F(y,c)=F(y',c)$ implies $x^L y = x^L y' \bmod p(x)$ which implies $y=y'$.
 Moreover, if we forbid the character value zero at the beginning of strings, and pick the initial value randomly, we have that 
 \textsc{Division} is uniform and thus, $\frac{nL}{2^{L}-2^{L/2+1}}$-almost strongly universal.

We might be able to compute \textsc{Division}  faster
than \textsc{CWPoly}. However, selecting a random irreducible polynomial might be slow.
To cope with this problem, Shoup introduced a generalized \textsc{Division}~\cite{shoup1996fast} with compression function $F(y,c)=y x^{L/k} + c $ and $p(x)$ chosen as a monic irreducible polynomial of degree $L/k$.

\subsection{Bernstein hashing}

Bernstein proposed a computationally efficient compression
function~\cite{bernstein22cdb}: $F(y,c)=  ( (y\ll l) + y) \oplus c$ where
$y\ll l$ is the left shift by $l$~bits. For all $l>0$, this compression function
is strongly permuting. 
Hence, given randomly chosen initial values, we have uniform hashing.

\subsection{Fowler-Noll-Vo hashing}

There are two types of Fowler-Noll-Vo  hash functions~\cite{fnvhash}.
The FNV-1 compression functions takes the form $F(y,c)= (y p) \oplus c$  where
$p$ is a prime number. It is a generalization of  Bernstein hashing. The FNV-1a compression functions are of the form
$F(y,c)=(y \oplus c) p$ for some prime $p$. Both FNV-1 and FNV-1a are strongly permuting.

\subsection{SAX and SXX}

The shift-add-xor (SAX) scheme~\cite{ramakrishna1997performance}
is defined by the compression function $F(y,c)=  y \oplus ( (y\ll l)+ (y\gg r)+ c)$
where  $y\gg r$ is the  right shift by $r$~bits.
For $L=32$ and 7-bit characters, Ramakrishna and Zobel~\cite{ramakrishna1997performance} 
reported that SAX is \emph{empirically universal} for $4\leq l \leq 7$ and $1\leq r \leq 3$, and a randomly chosen 32-bit initial value. They found
that the 
alternative, shift-xor-xor (SXX), $F(y,c)= y \oplus ( (y\ll l)\oplus(y\gg r)\oplus c)$, is not
competitive.

\subsection{String hashing functions in common programming languages}

Strings are commonly used as keys in hash tables.
Thus, most programming languages include 
string hashing functions. We consider C++ and Java.

ISO added support for hash tables to the C++ language (\texttt{unordered\_map})~\cite{cpp0x}.
Implementations of the language are required to provide a string hashing function, but the
exact function is unspecified. However, a popular compiler (GNU~GCC, version~4.1.1) implemented
it as an iterated hash function with the compression function $F(y,c)= 5 y+c \bmod{2^{32}}$ and an initial value of zero. For example, the hash value of the one-character string \texttt{z} is 122, the decimal value corresponding
to the character \texttt{z}.

The Java \texttt{String} class has a specified \texttt{hashCode} method. As of
version 1.3 of the language, it is an iterated hash
function with compression function $F(y,c)=31 y+c$---using \texttt{int} arithmetic---and an initial value of zero. Because Java lacks unsigned integers as a native type, 
the hash value of a sufficiently long string (e.g., \texttt{zzzzzz}) can be a negative integer.
(Java uses the Two's complement binary representation, so that signed integers are interchangeable with unsigned integers
as long as we only use addition,  subtraction and multiplication.)

\subsection{{\textsc{PowerOfTwo}} hashing}

In light of the hash functions used in Java and C++, consider the family given by the compression function $F(y,c)= B y +c \bmod{2^L}$ (henceforth \textsc{PowerOfTwo}). If the initial value is zero, then $h(00)=h(0)=0$, but we can fix this problem 
by using a non-zero initial value.
However, suppose that $B$ is even and consider any two strings $s$ and $s'$ of length greater than $L$ and differing only
in the first character, then $h(s)=h(s')$ because $B^L \bmod{2^L}=0$.
Thus---unsurprisingly---both the C++  and Java implementations set $B$ to an odd integer.

Suppose that $B$ is odd. The compression function is then strongly permuting. 
Thus, by choosing the initial value at random, we have uniform hashing.

When $B$ is odd, we have
that $B+1$ is even, so that $(B+1)^L \mod{2^L}=0$.
By the binomial theorem, we have
that $0=(B+1)^L \mod{2^L}= \sum_{k=0}^L B^k ( {L \choose k} \mod{2^L}) \mod{2^L}$.
Thus---irrespective of the initial value---the two strings of length $L+1$ given by the characters ${L \choose k} \mod{2^L}$ for $k=0,1,\ldots, L$ and $00\cdots 0$ collide when $B$ is odd. 
By this construction, 
\textsc{PowerOfTwo} cannot be almost universal unless we limit the length of the strings to at most $L$~characters. 

\end{document}